\documentclass[aip,jcp,10pt,twocolumn]{revtex4-1}
\usepackage{mathtools}
\usepackage{bbold}
\usepackage{amssymb}
\usepackage{subcaption}
\usepackage{multirow}
\usepackage{makecell}
\usepackage{booktabs}
\usepackage{xcolor}

\usepackage[normalem]{ulem} 

\usepackage{CJK} 

\newcommand\numberthis{\addtocounter{equation}{1}\tag{\theequation}}
\newcommand\mb{\mathbf}
\newcommand\mbs{\boldsymbol}
\newcommand\Id{\mbs{\mathbb{1}}}
\newcommand\Ov{\mathcal{O}}

\begin{document}

\begin{CJK*}{GB}{} 

\title{Ratchet-induced variations in bulk states of an active ideal gas}

\author{Jeroen Rodenburg}
\email{a.j.rodenburg@uu.nl}
\affiliation{Institute for Theoretical Physics, Utrecht University, Princetonplein 5, 3584 CC Utrecht, The Netherlands} 
\author{Siddharth Paliwal}
\affiliation{Soft Condensed Matter Group, Debye Institute for Nanomaterials Science, Utrecht University, Princetonplein 5, 3584 CC Utrecht, The Netherlands}
\author{Marjolein de Jager}
\affiliation{Institute for Theoretical Physics, Utrecht University, Princetonplein 5, 3584 CC Utrecht, The Netherlands} 
\author{Peter G. Bolhuis}
\affiliation{Van 't Hoff Institute for Molecular Sciences, University of Amsterdam, P.O. Box 94157, 1090 GD Amsterdam, The Netherlands}
\author{Marjolein Dijkstra}
\affiliation{Soft Condensed Matter Group, Debye Institute for Nanomaterials Science, Utrecht University, Princetonplein 5, 3584 CC Utrecht, The Netherlands}
\author{Ren\'e van Roij}
\email{r.vanroij@uu.nl}
\affiliation{Institute for Theoretical Physics, Utrecht University, Princetonplein 5, 3584 CC Utrecht, The Netherlands} 

\begin{abstract}
We study the distribution of active, noninteracting particles over two bulk states separated by a ratchet potential. By solving the steady-state Smoluchowski equations in a flux-free setting, we show that the ratchet potential affects the distribution of particles over the bulks, and thus exerts an influence of infinitely long range. As we show, crucial for having such a long-range influence is an external potential that is nonlinear. We characterize how the difference in bulk densities depends on activity and on the ratchet potential, and we identify power law dependencies on system parameters in several limiting cases. While weakly active systems are often understood in terms of an effective temperature, we present an analytical solution that explicitly shows that this is not possible in the current setting. Instead, we rationalize our results by a simple transition state model, that presumes particles to cross the potential barrier by Arrhenius rates modified for activity. While this model does not quantitatively describe the difference in bulk densities for feasible parameter values, it does reproduce - in its regime of applicability - the complete power law behavior correctly.

\end{abstract}
\maketitle

\end{CJK*} 

\section{Introduction}

\noindent Over the last few years, active matter has emerged as a testing ground for nonequilibrium statistical physics \cite{Seifert2016,ActiveEntropyProduction2,Tailleur2018,BradyReview,DDFTMicroswimmers,MCTActive,HermingHaus2016,PFTActive}. Its relevance comes from the fact that experimental realizations exist \cite{MicroSwimmersReview,PoonEcoli,Drescher} of relatively simple active matter models, such as active Brownian particles (ABPs) and run-and-tumble (RnT) particles \cite{ABPvsRnT}. While describing these systems can be very challenging when they are far from thermodynamic equilibrium \cite{TailleurGeneralThermo,Siddharth2018}, for small activity they are well understood by effective equilibrium approaches \cite{Brader2015,SpeckEffectivePairPotential,MarconiMaggi,MarconiMaggiEffectivePotential,Binder2016}. In particular, it is well established that noninteracting particles at small activity can be described as an equilibrium system at an effective temperature \cite{CugliandoloEffTemp,Wang2011,MarconiMaggi,Fily2012,Szamel2014}. For example, inserting the effective temperature in the Einstein relation yields the enhanced diffusion coefficient of an active particle, and using the effective temperature in the Boltzmann distribution gives the distribution of weakly active particles in a gravitational field \cite{Palacci2010,ES,EoSExperiment,Wolff2013,Szamel2014,ABPvsRnT,EoSExperiment,Stark2016}.\\
\begin{figure}
 \centering
  \includegraphics[width=\linewidth]{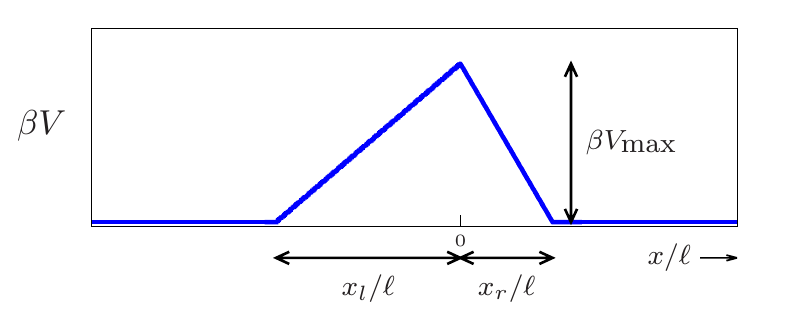}
 \caption{\raggedright (Dimensionless) ratchet potential $\beta V$, as a function of the Cartesian $x$-coordinate in units of the diffusive length scale $\ell$. The ratchet can be characterized by its height $\beta V_{\text{max}}$, the width of its left side $x_l/\ell$, and its asymmetry $a = (x_l - x_r)/x_r$.}
  \label{fig:RatchetPotential}
\end{figure}
	\indent However, even weakly active systems can display behavior very different from equilibrium systems \cite{YaouensTip,Galajda2007, ActiveRatchets2011,RatchetEffectsReview,Reichhardt2016CollectiveRatchet,Nikola2016,UnderdampedRatchet2017, Caleb2017,Yongjoo2018}. For instance, a single array of funnel-shaped barriers, that is more easily crossed from one lateral direction than from the other, can induce a steady state with ratchet currents  that span the entire system \cite{RatchetEffectsReview}. Alternatively, when the boundary conditions deny such a system-wide flux, the result is a steady state with a higher density on one side of the array than on the other \cite{Galajda2007,RatchetEffectsReview}. As the system can be arbitrarily long in the lateral direction, the presence of the funnels influences the density profile at arbitrarily large distance.\\
	\indent Needless to say, characterizing such a long-range effect is a challenge, and the natural place to start is in a setting as simple as possible. As we shall show, having an external potential with a long-range influence on the density profile in steady state is only possible with the key ingredients of (1) activity, and (2) an external potential that is nonlinear. Therefore, a good candidate for a minimal model is to study the distribution of active particles over two bulks separated by a potential barrier that is only piecewise linear. Here, we focus on a sawtooth-shaped barrier, known as a ratchet potential (see Fig. \ref{fig:RatchetPotential}). As we will see, the asymmetry of the ratchet induces a flux-free steady state with different densities in both bulks. Since the bulk sizes can be arbitrarily large, the influence of the ratchet potential is indeed of infinite range. This system has actually already been studied, both experimentally \cite{DiLeonardo2013} and theoretically \cite{ActiveRatchets2014}. However, the former study was performed at high degree of activity, and the latter study neglected Brownian fluctuations, such that the degree of activity could not be quantified. Thereby, the regime of weak activity, where the statistical physics generally seems best understood \cite{Brader2015,SpeckEffectivePairPotential,MarconiMaggi,MarconiMaggiEffectivePotential,Binder2016,Siddharth2018}, remains largely unexplored.\\
	\indent In this work, we study the effect of an external potential on arbitrarily large bulk regions with as few complications as possible. To this end, we investigate how a ratchet potential affects active particles that also undergo translational Brownian motion, such that the degree of activity can be quantified. We ask the questions: can we characterize how the external potential influences the density distribution as a function of activity? And can we understand this distribution in the limit of weak activity?\\

\noindent The article is organized as follows. In section \ref{sec:Models}, we introduce two active particle models, as well as the ratchet potential. In section \ref{sec:NumSol}, we numerically solve the density and polarization profiles of these active particles in the ratchet potential, and we study how the difference in bulk densities depends on activity, and on the ratchet potential. In section \ref{sec:SmallfpSol}, we specialize to the limit of weak activity, and provide an analytical solution that explicitly shows that the nonzero difference in bulk densities cannot be understood by the use of an effective temperature. Instead, in section \ref{sec:TSModel}, we propose to understand the density difference in terms of a simple transition state model. We end with a discussion, in section \ref{sec:Discussion}, on what ingredients are necessary to have the external potential affect the densities in such a (highly) nonlocal way, and with concluding remarks in section \ref{sec:Conclusion}.

%
%
%
%
%
%
\section{Models}
\label{sec:Models}
\subsection{2D ABPs}
\noindent In order to investigate the behavior of active particles in a ratchet potential, we consider the widely employed model of active Brownian particles\cite{ABPs} (ABPs) in two dimensions. For simplicity, we consider spherical, noninteracting particles. Every particle is represented by its position $\mb{r}(t) = x(t) \mb{\hat{x}} + y(t) \mb{\hat{y}}$, where $\mb{\hat{x}}$ and $\mb{\hat{y}}$ are Cartesian unit vectors and $t$ is time, as well as by its orientation $\mb{\hat{e}}(t) \equiv \cos\theta(t) \mb{\hat{x}} + \sin\theta(t) \mb{\hat{y}}$. Its time evolution is governed by the overdamped Langevin equations
\begin{subequations}
 \label{eqn:Langevin}
\begin{align*}
  \partial_t\mb{r}(t)
  &=
  v_0 \mb{\hat{e}}(t) 
  - \gamma^{-1}\mbs{\nabla}V(\mb{r})
  + \sqrt{2D_t} \mbs{\eta}_t(t),
 \label{eqn:Langevina} \numberthis\\
  \partial_t \theta(t) 
  &= 
  \sqrt{2D_r} \eta_r(t).
 \label{eqn:Langevinb}\numberthis
\end{align*}
\end{subequations}
\noindent Eq. (\ref{eqn:Langevina}) expresses that a particle's position changes in response to (i) a propulsion force, that acts in the direction of $\mb{\hat{e}}$, and that gives rise to a propulsion speed $v_0$, (ii) an external force, generated by the external potential $V(\mb{r})$, and (iii) the unit-variance Wiener process $\mbs{\eta}_t(t)$, that gives rise to translational diffusion with diffusion coefficient $D_t$. Here $\gamma$ is the friction coefficient. Note that $\beta \equiv (\gamma D_t)^{-1}$ is an inverse energy scale, and that in thermodynamic equilibrium the Einstein relation implies $\beta = (k_BT)^{-1}$, where $k_B$ is the Boltzmann constant and $T$ the temperature. Eq. (\ref{eqn:Langevinb}) expresses that the orientation of a particle changes due to the unit-variance Wiener process $\eta_r(t)$, which leads to rotational diffusion with diffusion coefficient $D_r$.\\
	\indent The stochastic Langevin equations (\ref{eqn:Langevin}) induce a probability density $\psi(\mb{r},\theta,t)$, whose time evolution follows the Smoluchowski equation
\begin{flalign*}
 \label{eqn:SE}
  \partial_t\psi
  =
  -\mbs{\nabla} \cdot
    \Big( 
     v_0 \mb{\hat{e}}\psi
     - \frac{1}{\gamma} (\mbs{\nabla}V) \psi
     - D_t \mbs{\nabla} \psi
    \Big) 
   \mathrlap{+ D_r \partial_{\theta\theta} \psi.}
 \numberthis &&
\end{flalign*}
\noindent Here $\mbs{\nabla} = (\partial_x,\partial_y)^T$ is the two-dimensional spatial gradient operator. 
Two useful functions to characterize the probability density $\psi(\mb{r},\theta,t)$ are the
density $\rho(\mb{r},t) \equiv \int\mathrm{d}\theta\psi(\mb{r},\theta,t)$
and the polarization
$\mb{m}(\mb{r},t) \equiv \int\mathrm{d}\theta\psi(\mb{r},\theta,t)\mb{\hat{e}}(\theta)$.
\noindent Their time-evolutions follow from the Smoluchowski equation (\ref{eqn:SE}) as
\begin{align*}
 \label{eqn:SEMoments}
 \partial_t \rho
 &=
 -\mbs{\nabla} \cdot 
  \Big\{
   v_0 \mb{m}   
   -\frac{1}{\gamma}(\mbs{\nabla}V)\rho
   -D_t\mbs{\nabla}\rho
  \Big\},
 \numberthis \\
 \partial_t\mb{m}
 &=
 - \mbs{\nabla} \cdot
  \Big\{
   v_0 \big(\mbs{\mathcal{S}} + \frac{\Id}{2}\rho\big)   
   - \frac{1}{\gamma}(\mbs{\nabla}V)\mb{m}
   - D_t \mbs{\nabla} \mb{m}
  \Big\}
  -D_r\mb{m},
\end{align*}
\noindent where $\Id$ is the $2 \times 2$ identity matrix, and where $\mbs{\mathcal{S}}(\mb{r},t) \equiv \int\mathrm{d}\theta\psi(\mb{r},\theta,t)(\mb{\hat{e}}(\theta)\mb{\hat{e}}(\theta)-\Id/2)$ is the $2 \times 2$ nematic alignment tensor. Due to the appearance of $\mbs{\mathcal{S}}$, Eqs. (\ref{eqn:SEMoments}) are not closed. Therefore, solving Eqs. (\ref{eqn:SEMoments}), rather than the full Smoluchowski Eq. (\ref{eqn:SE}), requires a closure, an example of which we discuss in section \ref{subsec:1DRnT}.\\
	\indent We consider a planar geometry that is invariant in the $y$-direction, i.e. $V(\mb{r}) = V(x)$, such that $\psi(\mb{r},\theta,t)=\psi(x,\theta,t)$, $\rho(\mb{r},t) = \rho(x,t)$, $\mb{m}(\mb{r},t) = m_x(x,t)\mb{\hat{x}}$ etc. The geometry consists of two bulks, located at $x\ll0$ and $x\gg0$. These bulk systems are separated by the ratchet potential
\begin{flalign*}
 \label{eqn:RatchetPotential}
  V(x)
  =
  \left\{
   \begin{alignedat}{2}
    &\quad0, \quad&&\text{for}\quad x<-x_l,\\
    &V_{\text{max}}\Big(\frac{x}{x_l}+1\Big), \quad&&\mathrlap{\text{for}\quad -x_l < x < 0,}\\
    &V_{\text{max}}\Big(1-\frac{x}{x_r}\Big), \quad&&\mathrlap{\text{for}\quad 0 < x < x_r,}\\
    &\quad0, \quad&&\text{for}\quad x_r < x,
   \end{alignedat}
  \right.
 \numberthis&&
\end{flalign*}
\noindent where $x_l$ and $x_r$ are both positive. This sawtooth-shaped potential is illustrated in Fig. \ref{fig:RatchetPotential}. Note that the potential is generally asymmetric, the degree of which is characterized by the asymmetry factor $a \equiv (x_l-x_r)/x_r$. Without loss of generality, we only consider ratchets for which $x_l > x_r$, such that $a>0$.

	\indent The complete problem is specified by four dimensionless parameters. We use the rotational time $D_r^{-1}$, and the diffusive length scale $\ell \equiv \sqrt{D_t/D_r}$, which is proportional to the size of a particle undergoing free translational and rotational diffusion, to obtain the Peclet number
\begin{flalign*}
 \begin{alignedat}{2}
  &\text{Pe} \equiv \frac{1}{\sqrt{2}}\frac{v_0}{D_r \ell}, &&\text{ as a measure for the degree of } \mathrlap{\text{activity,}}\\
  &\beta V_{\text{max}}, &&\text{ the barrier height},\\
  &\frac{x_l}{\ell}, &&\text{ the width of the ratchet's left side,}\\
  &a, &&\text{ the asymmetry of the ratchet}.
 \end{alignedat}
\label{eqn:DimlessPars}
\numberthis &&
\end{flalign*}
\noindent We caution the reader that the factor $1/\sqrt{2}$ is often omitted from the definition of the Peclet number; it is included here to connect to the model described below.
\subsection{1D RnT}
\label{subsec:1DRnT}
\noindent The fact that there is only one nontrivial dimension in the problem suggests a simpler, one-dimensional model with the same physical ingredients. In this model, which we refer to as the 1D Run and Tumble (RnT) model, particles are characterized by a position $x(t)$, as well as by an orientation $e_x(t)$ that points in either the positive or the negative $x$-direction, i.e. $e_x = \pm 1$. The orientation $e_x$ can flip with probability $D_r$ per unit time. Every particle performs overdamped motion driven by (i) a propulsion force, that acts in the direction of its orientation, (ii) an external force, generated by the ratchet potential (\ref{eqn:RatchetPotential}), and (iii) Brownian motion, with associated diffusion constant $D_t$. The problem can be specified in terms of probability density functions $\psi_{\pm}(x,t)$ to find particles with orientation $e_x = \pm 1$. For our purposes, it is more convenient to consider the density $\rho(x,t) \equiv \psi_+(x,t) + \psi_-(x,t)$, and polarization $m_x(x,t) \equiv [\psi_+(x,t) - \psi_-(x,t)]/\sqrt{2}$. These fields evolve as
\begin{flalign*}
 \label{eqn:RnTEqs}
 \begin{alignedat}{1}
    \partial_t \rho
   &=
   -\partial_x
   \Big\{    
    \sqrt{2} v_0 m_x   
    -\frac{1}{\gamma}(\partial_x V)\rho
    -D_t\partial_x\rho 
   \Big\} 
  ,\\
   \partial_tm_x
   &=
   - \partial_x
   \Big\{
    \frac{v_0}{\sqrt{2}} \rho
    - \frac{1}{\gamma}(\partial_xV)m_x
    - D_t \partial_x m_x
   \Big\}
    \mathrlap{-D_r m_x.}
 \end{alignedat}
 \numberthis &&
\end{flalign*}
\noindent Note the similarity of Eqs. (\ref{eqn:RnTEqs}) with Eqs. (\ref{eqn:SEMoments}) of the 2D ABP model. In fact, if we define the Peclet number for the 1D RnT model as $\text{Pe} \equiv v_0/(D_r \ell)$, then supplying the 2D ABP model with the closure $\mbs{\mathcal{S}}(\mb{r},t) = 0$ maps Eqs. (\ref{eqn:SEMoments}) to the 1D RnT model. The mapping is such that if one uses the same values for the dimensionless parameters $\text{Pe}, \beta V_{\text{max}}$, $x_l/\ell$, and $a$, then both models yield equal density profiles $\rho(x)$ and polarization profiles $m_x(x,t)$.
As the closure $\mbs{\mathcal{S}}(\mb{r},t) = 0$ is exact in the limit of weak activity, i.e. $\text{Pe} \ll 1$, this mapping is expected to give good agreement between the two models for small values of the Peclet number $\text{Pe}$.

\section{Numerical solutions}
\label{sec:NumSol}
\subsection{Density and mean orientation profiles}
\label{subsec:NumSolA}
\begin{figure}
  \includegraphics[width=\linewidth]{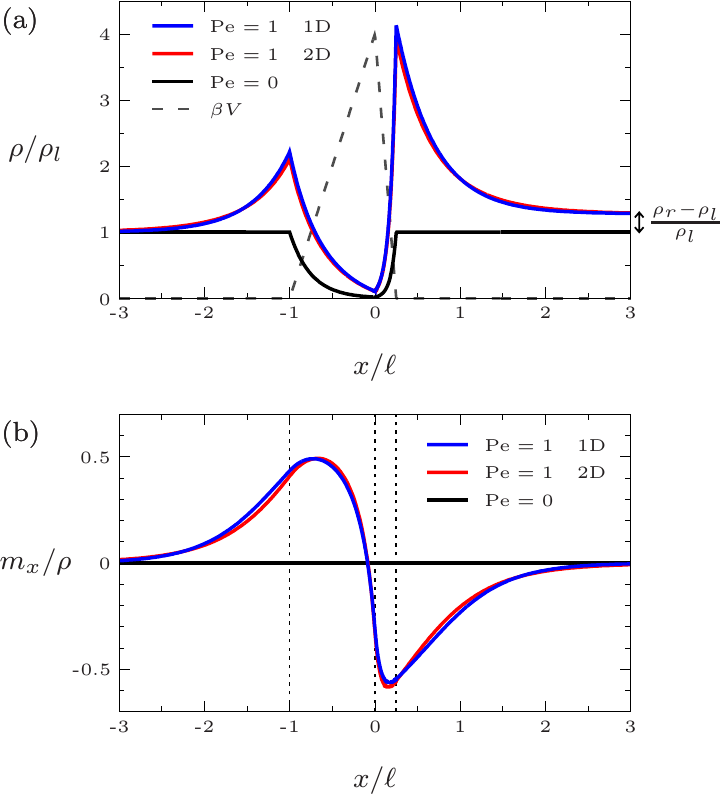}
 \caption{\raggedright (a) Density profiles $\rho(x)/\rho_l$ and (b) mean orientation profiles $m_x(x)/\rho(x)$ of 2D ABPs, and 1D RnT particles, as indicated, in a ratchet potential $V(x)$ of height $\beta V_{\text{max}} = 4$, width $x_l/\ell = 1$, and asymmetry $a=3$. The dashed, vertical lines indicate the positions of the barrier peak ($x=0$) and the ratchet sides ($x=-x_l$ and $x=x_r)$. The bulk density to the left of the ratchet is $\rho(x \ll -x_l) = \rho_l$. Passive particles ($\text{Pe} = 0$) are distributed isotropically ($m_x = 0$), with a density profile given by the Boltzmann weight $\rho(x) = \rho_l \exp(-\beta V(x))$. Consequently, the densities $\rho_l$ and $\rho_r$ in the bulks on either side of the ratchet are equal. Active particles ($\text{Pe} = 1$) display much richer behaviour, with an accumulation of particles at either side of the ratchet, with a mean orientation towards the barrier peak, with a depletion of particles near the top of the ratchet, and with the right bulk density $\rho_r$ exceeding the left bulk density $\rho_l$.} 
 \label{fig:TypicalProfiles}
\end{figure}
\noindent We study steady state solutions of both 2D ABPs and 1D RnT particles in the ratchet potential (\ref{eqn:RatchetPotential}). To find the solutions, for the 2D ABP model we numerically solve Eq. (\ref{eqn:SE}) with $\partial_t\psi = 0$, whereas for the 1D model we numerically solve Eqs. (\ref{eqn:RnTEqs}) with $\partial_t \rho = \partial_t m_x = 0$. We impose the following three boundary conditions. 
\begin{enumerate}
 \item{To the left of the ratchet, we imagine an infinitely large reservoir that fixes the density to be $\rho_l$ at $x_{\text{res}} \ll -x_l$, i.e. we impose $\psi(x_{\text{res}},\theta) = (2\pi)^{-1} \rho_l$ for the 2D case, and $\rho(x_{\text{res}}) = \rho_l$, $m_x(x_{\text{res}}) = 0$ for the 1D case.}
 \item{To the right of the ratchet, we assume an isotropic bulk that is thermodynamically large, yet finite, such that its density follows from the solution of the equations. In technical terms, at $x_{\text{max}} \gg x_r$ we impose $\partial_x \psi(x_{\text{max}},\theta) = 0$ for the 2D case, and $\partial_x \rho(x_{\text{max}})=0$, $m_x(x_{\text{max}}) = 0$ for the 1D case.}
 \item{Additionally, for the 2D case we assume periodic boundary conditions, i.e. $\psi(x,0)=\psi(x,2\pi)$ and $\partial_{\theta}\psi(x,0)=\partial_{\theta}\psi(x,2\pi)$ for all $x$.}
\end{enumerate}

\noindent In order to allow the profiles to decay to their bulk values specified by boundary conditions 1 and 2, in our numerical calculations we always ensure the distance between $x_\text{res}$ (or $x_\text{max}$) and the ratchet potential to be at least a multitude of the most significant length scale.\\
	\indent Typical solutions are shown in Fig. \ref{fig:TypicalProfiles}. The considered ratchet potential, with height $\beta V_{\text{max}} = 4$, width $x_l / \ell = 1$, and asymmetry $a=3$, is shown as the dashed line in Fig. \ref{fig:TypicalProfiles}(a). We consider both a passive system ($\text{Pe} = 0$) and an active system ($\text{Pe} = 1$), using $x_\text{res}=-11l$ and $x_\text{max}=10.25l$ in this case. The resulting density profiles and mean orientation profiles are shown in Fig. \ref{fig:TypicalProfiles}(a) and Fig. \ref{fig:TypicalProfiles}(b), respectively. For the passive system, the solution is isotropic (i.e. $\psi(x,\theta) \propto \rho(x)$ and $m_x(x) = 0$ everywhere), and given by the Boltzmann weight $\rho(x) = \rho_l\exp(-\beta V(x))$. One checks that these solutions indeed solve Eqs. (\ref{eqn:SE}) and (\ref{eqn:RnTEqs}) when the propulsion speed $v_0$ equals $0$. Thus, in accordance with this Boltzmann distribution, the density in the passive system is lower in the ratchet region than in the left bulk, and its value $\rho_r \equiv \rho(x_{\text{max}})$ in the right bulk satisfies $\rho_r = \rho_l$, with $\rho_l$ the density in the left bulk. This is a necessity in thermodynamic equilibrium, even for interacting systems: the equality of the external potential implies equal densities of the bulks.\\
	\indent For the active case ($\text{Pe} = 1$), the behavior is much richer. Firstly, the solution is anisotropic in the ratchet region, even though the external potential is isotropic. Indeed, Fig. \ref{fig:TypicalProfiles}(b) shows a mean orientation of particles directed towards the barrier on either side of the ratchet. This is consistent with the finding that active particles tend to align against a constant external force \cite{ES,Sedimentation2018}, but is also reminiscent of active particles near a repulsive wall. Indeed, at walls particles tend to accumulate with a mean orientation towards the wall \cite{GompperWallAccumulation,Ran2015}, and a similar accumulation is displayed by the density profiles of Fig. \ref{fig:TypicalProfiles}(a) at the ratchet sides $x = -x_l$ and $x = x_r$. The overall result is an accumulation of particles at the ratchet sides, a depletion of particles near the center of the ratchet, and, remarkably, a density $\rho_r$ in the right bulk that is higher than the density $\rho_l$ in the left bulk.\\
	\indent The fact that the difference in bulk densities $\Delta \rho \equiv \rho_r - \rho_l$ is positive is caused by the asymmetry of the ratchet: due to their propulsion force, particles can cross the potential barrier more easily from the shallower, left side than from the steeper, right side. This argument is easily understood in the absence of translational Brownian motion ($D_t = 0$), i.e. when the only force that makes particles move (apart from the external force) is the propulsion force. Indeed, in this case, one can even think of ratchet potentials whose asymmetry is such that particles \emph{can} climb it from the shallow side, but \emph{not} from the steep side \cite{ActiveRatchets2014}. For such a ratchet potential, \emph{all} particles eventually end up on the right side of the ratchet, such that clearly the right bulk density $\rho_r$ exceeds the left bulk density $\rho_l$. The effect of having nonzero translational Brownian motion ($D_t > 0$) is that particles always have \emph{some} probability to climb also the steep side of the ratchet. This leads to a density difference $\Delta \rho$ that is smaller than in the $D_t = 0$ case. Yet, as long as the ratchet is asymmetric, the density difference \emph{always} turns out positive for \emph{any} positive activity $\text{Pe}$.\\
	\indent We stress that the fact that $\rho_r > \rho_l$ is actually quite remarkable. The reason is that, whereas the ratchet potential is localized around $x=0$, the right bulk can be arbitrarily large. Since our results clearly show that the right bulk density $\rho_r$ is influenced by the ratchet, this means that the range of influence of the external potential is in some sense infinitely large. 
\\
\begin{figure*}[t]
  \includegraphics[width=\linewidth]{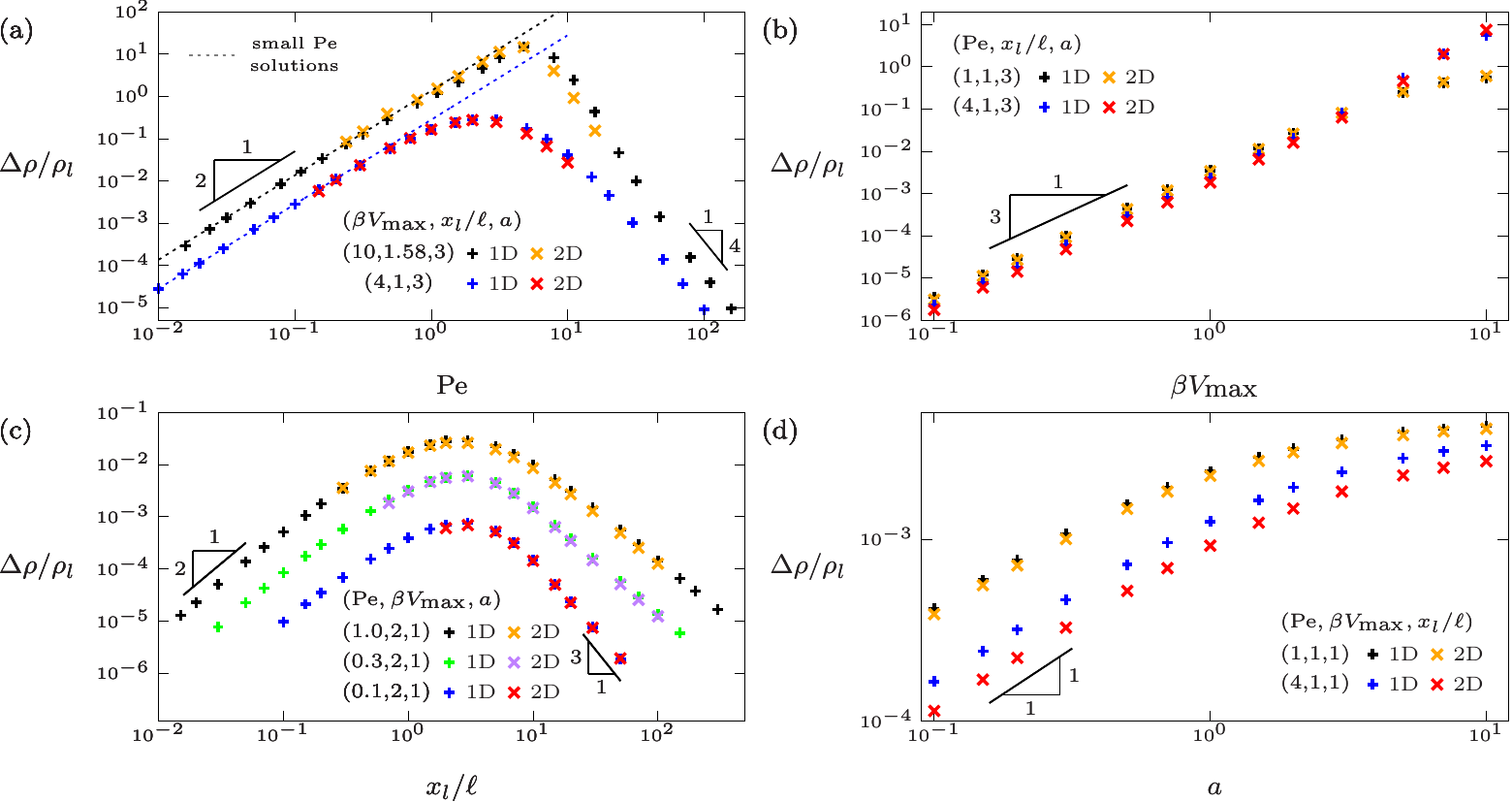}
  \caption{\raggedright Normalized density difference $\Delta \rho/\rho_l$ as a function of (a) activity $\text{Pe}$, (b) barrier height $\beta V_{\text{max}}$, (c) barrier width $x_l/\ell$, and (d) barrier asymmetry $a$. Results are shown for both the 2D ABP and 1D RnT models, as indicated. In the limiting cases of small and large values of its arguments, the density difference shows power law behavior. The corresponding exponents are listed in Table \ref{tab:powerlaws}. Additionally, (a) shows the density difference obtained analytically in the limit of weak activity (see section \ref{sec:SmallfpSol}), for the same ratchet parameters as used for the numerical solutions. The analytical and numerical solutions show good agreement up to $\text{Pe} \approx 0.5$.}
 \label{fig:DensityDifference}
\end{figure*}
\begin{table}
 \begin{ruledtabular}
 \begin{tabular}{@{}c c c c c@{}}
   & & \multicolumn{3}{c}{exponent} \\\cline{3-5}
  base & limit & \makecell{numerical \\ solution} & \makecell{Pe $\ll 1$ \\ solution} & \makecell{transition state \\ model} \\\colrule
  Pe & Pe $\ll 1$ & 2 & 2 & 2 \\
                      & Pe $\gg 1$ & -4 & & \\
  $\beta V_{\text{max}}$ & $\beta V_{\text{max}} \ll 1$ & 3 & 3 &   \\
                         & $\beta V_{\text{max}} \gg 1$ & 0 & 0 & 0 \\
  $x_l/\ell$ & $x_l/\ell \ll 1$ &  2  & 2  & 2 \\
             & $x_l/\ell \gg 1$ &  * & -3 &   \\
  $a$ & $a \ll 1$ & 1 & 1 & 1 \\
      & $a \gg 1$ & 0 & 0 & 0
 \end{tabular}
 \end{ruledtabular}
 * depends on Pe. For $\text{Pe} \ll 1$, this exponent equals $-3$.
 \caption{\raggedright Power laws $\Delta \rho \propto \text{base}^{\text{exponent}}$, for limiting values of the base. Here the base denotes either the activity Pe, the barrier height $\beta V_{\text{max}}$, the barrier width $x_l/\ell$, or the barrier asymmetry $a$. Exponents were obtained numerically for the 1D RnT and 2D ABP models (yielding consistent exponents), analytically for the case of small activity $\text{Pe} \ll 1$, and for a simple transition state model. Exponents are shown only in limits where the corresponding solution is applicable.}
 \label{tab:powerlaws}
\end{table}
\subsection{Scaling of the bulk density difference $\Delta \rho$}
\label{subsec:NumSolB}
\noindent Next, we examine, one by one, how the density difference $\Delta\rho$ depends on activity $\text{Pe}$, the barrier height $\beta V_{\text{max}}$, the barrier width $x_l/\ell$, and on the barrier asymmetry $a$. The results are shown in Figs. \ref{fig:DensityDifference}(a)-(d), for both the 2D ABP and the 1D RnT models. In all cases, both models give density differences that are quantitatively somewhat different, but qualitatively similar, as they are both consistent with identical power laws\footnote{For numerical reasons, fewer results were obtained for the 2D ABP model than for the 1D RnT model. Therefore, not all of the power laws obtained for the 1D model could be tested for the 2D model. Yet, all 2D results seem consistent with all of the power laws.}.\\
	\indent Fig. \ref{fig:DensityDifference}(a) shows the density difference as a function of activity $\text{Pe}$, for two different ratchet potentials. For small $\text{Pe}$, the figure shows that the density difference increases as $\text{Pe}^2$. For large $\text{Pe}$, the density difference decreases again, to decay to $0$ in the limit $\text{Pe} \rightarrow \infty$. The reason for this decrease is that particles with high activity can easily climb either side of the ratchet potential, such that they hardly notice the presence of the barrier at all. As shown by Fig. \ref{fig:DensityDifference}(a), this decay follows the power law $\Delta \rho \propto \text{Pe}^{-4}$. Whereas the prefactors of these power laws are different for the two different ratchet potentials considered, the exponents were found to be independent of the ratchet parameters, which was tested for many more values of $\beta V_{\text{max}}$, $x_l/\ell$, and $a$.\\ 
	\indent Fig. \ref{fig:DensityDifference}(b) shows the density difference as a function of the barrier height $\beta V_{\text{max}}$. The barrier width, $x_l/\ell = 1$, and asymmetry, $a=3$, are kept fixed, and two levels of activity, $\text{Pe} = 1$ and $\text{Pe}=4$, are considered. For all cases, we find the power law $\Delta \rho \propto (\beta V_{\text{max}})^3$, up to values of the barrier height $\beta V_{\text{max}} \approx 3$. Exploring the behavior for large values of the barrier height $\beta V_{\text{max}}$ was numerically not feasible, but the fact that the curves for activity $\text{Pe} = 1$ level off for barrier heights $\beta V_{\text{max}} \geq 5$ seems consistent with the asymptotic behavior for $\beta V_{\text{max}} \gg 1$  that we shall obtain, in section \ref{sec:SmallfpSol}, in the limit of weak activity.\\
	\indent Fig. \ref{fig:DensityDifference}(c) shows the density difference as a function of the width $x_l/\ell$ of the left side of the ratchet. Here the barrier height and asymmetry are fixed, at $\beta V_{\text{max}} = 2$ and $a=1$, respectively, whereas the degree of activity is varied as $\text{Pe} = 0.1, 0.3,$ and $1$. For small barrier widths, i.e. for $x_l/\ell \ll 1$, the curves show the power law $\Delta \rho \propto (x_l/\ell)^{2}$, independent of the activity $\text{Pe}$. For very wide barriers, i.e. for $x_l/\ell \gg 1$, the curves show power law behavior with an exponent that does depend on the activity $\text{Pe}$. For the smallest degree of activity, $\text{Pe} = 0.1$, this exponent is found to equal $-3$. This scaling, $\Delta \rho \propto (x_l/\ell)^{-3}$ for large widths $x_l/\ell \gg 1$, will also be obtained analytically in section \ref{sec:SmallfpSol} for the case of weak activity.\\
	\indent Finally, Fig. \ref{fig:DensityDifference}(d) shows the density difference as a function of the barrier asymmetry $a$. The barrier height and width are fixed, at $\beta V_{\text{max}} = 1$ and $x_l/\ell = 1$, respectively, and the degree of activity is varied as $\text{Pe} = 1$ and $\text{Pe} = 4$. For nearly symmetric ratchets, i.e. for $a \ll 1$, all curves show $\Delta \rho \propto a$, whereas for large asymmetries $a \gg 1$ the curves suggest asymptotic behavior, i.e. $\Delta \rho \propto a^0$. This asymptotic behavior can be understood on physical grounds, as the limit $a\rightarrow \infty$ corresponds to a ratchet whose right slope is vertical, a situation that we expect to lead to a finite density difference indeed.\\
	\indent All discussed scalings are summarized in Table \ref{tab:powerlaws}. Of these, the scaling $\Delta \rho \propto \text{Pe}^{2}$ for small activity $\text{Pe} \ll 1$ can be regarded as trivial. The reason is that, in an expansion of the density difference $\Delta \rho$ around $\text{Pe} = 0$, the quadratic term is the first term to be expected on general grounds: (i) Eqs. (\ref{eqn:SE}) and (\ref{eqn:RnTEqs}) are invariant under a simultaneous inversion of the self-propulsion speed ($v_0 \rightarrow -v_0$) and the orientation ($\mb{\hat{e}} \rightarrow -\mb{\hat{e}}$, and hence $m_x \rightarrow -m_x$), such that the expansion of the density difference $\Delta \rho$ contains only even powers of $\text{Pe}$, and (ii) for the passive case ($\text{Pe} =0$), the density difference $\Delta \rho$ equals $0$, such that the zeroth order term is absent. Similarly, the obtained scaling $\Delta \rho \propto a$ is as expected: since a symmetric ratchet ($a=0$) leads to the density difference $\Delta \rho = 0$, the leading order term one expects in an expansion of the density difference $\Delta \rho$ around $a=0$ is linear in the asymmetry $a$. However, all other scalings listed in Table \ref{tab:powerlaws} cannot be predicted by such general arguments, and are therefore nontrivial findings.\\
	\indent We emphasize that these results have been obtained and verified by multiple approaches independently. While the presented results have been obtained by numerically solving the differential equations (\ref{eqn:SE}) and (\ref{eqn:RnTEqs}) as explained above, both the 2D ABP model and the 1D RnT model were also solved by separate approaches. For the 2D ABP model, results were additionally obtained by numerically integrating the Langevin equations (\ref{eqn:Langevin}) in particle-based computer simulations. For the 1D RnT model, results were also obtained by solving a lattice model, where particles can hop to neighbouring lattice sites, and change their orientation, with probabilities that reflect the same physical processes of self-propulsion, external forcing, translational Brownian motion, and tumbling\cite{MdJ}. For both the 2D ABP and the 1D RnT model, the two alternative approaches showed full agreement with the presented results.

\section{Weak activity limit}
\label{sec:SmallfpSol}
\noindent Having characterized how the ratchet potential influences the densities of the adjoining bulks, we now turn to the question whether we can better understand this effect. We first try to answer this question for the simplest case possible, and therefore focus on the limit of weak activity, i.e. $\text{Pe} \ll 1$. Recall that in this limit the 2D ABP model and the 1D RnT model are equivalent. In this section, we present an analytical solution for the $\text{Pe} \ll 1$ limit. In the next section, we propose to rationalize its results by a simple transition state model, that is valid for, but not limited to, weak activity.\\

\begin{figure}
 \includegraphics[width=\linewidth]{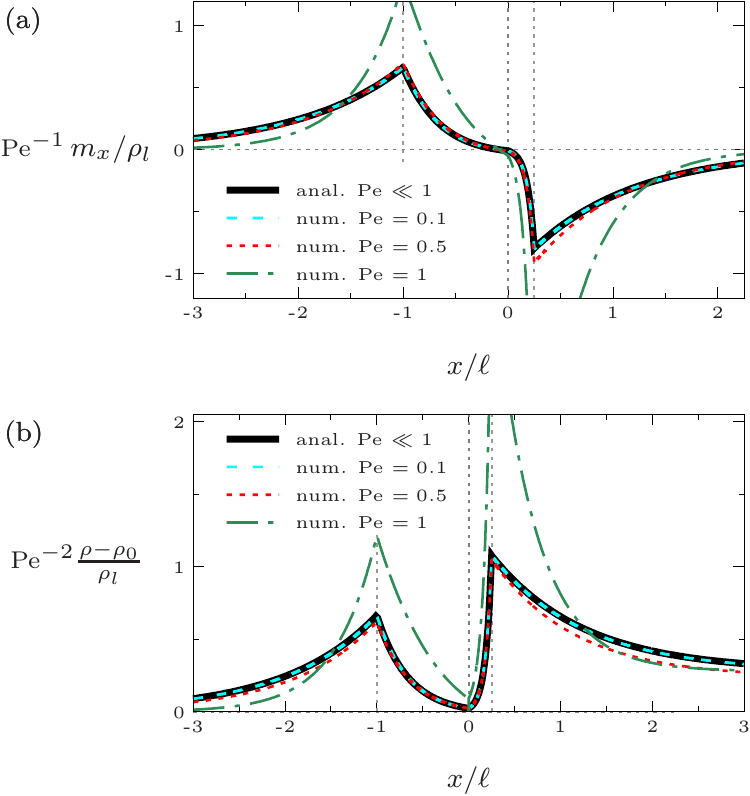}
 \caption{\raggedright (a) Normalized polarization profiles $m_x(x)/\rho_l$ and (b) deviations of the density $\rho(x)$ from the passive solution $\rho_0(x)$, for a ratchet potential of height $\beta V_{\text{max}} = 4$, width $x_l/\ell = 1$, and asymmetry $a=3$. The dashed, vertical lines indicate the positions of the barrier peak ($x=0$) and the ratchet sides ($x=-x_l$ and $x=x_r)$. Results are shown for the analytical Pe $\ll 1$ solution, and for the numerical solutions to the 1D RnT model, for activity levels Pe $=0.1, 0.5$ and $1$. The polarizations and density deviations are divided by Pe and Pe$^2$, respectively, such that the curves for the analytical solution are independent of Pe.}
 \label{fig:SmallfpPlots}
\end{figure}
\noindent In case of a small propulsion force, i.e. of $\text{Pe} \ll 1$, the density can be expanded as $\rho(x) = \rho_0(x) + \text{Pe}^2 \rho_2(x) + \Ov(\text{Pe}^4)$, and the polarization as $m_x(x) = \text{Pe} \, m_1(x) + \Ov(\text{Pe}^3)$. Here $\rho_0(x)$, $\rho_2(x)$ and $m_1(x)$ are assumed to be independent of $\text{Pe}$. We used the arguments that the density $\rho(x)$ is an even function of $\text{Pe}$, and the polarization $m_x(x)$ an odd function of $\text{Pe}$, as explained in section \ref{subsec:NumSolB}. With these expansions, Eqs. (\ref{eqn:RnTEqs}) can be solved perturbatively in $\text{Pe}$, separately for each region where the ratchet potential (\ref{eqn:RatchetPotential}) is defined. As shown in the appendix, the solutions within one region are
\begin{align*}
 \label{eqn:SmallfpSol}
  \rho_0(x) = &A_0 e^{-\beta V(x)},\\
  m_1(x) = &-\frac{A_0}{\sqrt{2}} f e^{-\beta V(x)}
            + B_+ e^{c_+x/\ell}
            + B_- e^{c_-x/\ell},\\
  \rho_2(x) = &\left[A_2-A_0 f \frac{x}{\ell} \right] e^{-\beta V(x)}\\
               &+ \frac{\sqrt{2}B_+}{c_+ - f} e^{c_+x/\ell}
                + \frac{\sqrt{2}B_-}{c_- - f} e^{c_-x/\ell}.
 \numberthis
\end{align*}
\noindent Here we defined the non-dimensionalized external force $f(x) \equiv -\beta \ell \partial_x V(x)$, such that $f=0$ for $x<-x_l$, $f = -\beta V_{\text{max}}\ell/x_l$ for $-x_l < x <0$, $f = \beta V_{\text{max}}\ell/x_r$ for $0<x<x_r$, and $f=0$ for $x>x_r$, in accordance with Eq. (\ref{eqn:RatchetPotential}). Furthermore, we defined $c_{\pm} \equiv (f\pm\sqrt{f^2+4})/2$. The integration constants $A_0, A_2, B_+,$ and $B_-$ are found separately for each region, by applying the boundary conditions $\rho(-\infty)=\rho_l$, $m(\infty)=m(-\infty)=0$, and the appropriate continuity conditions at the region boundaries $x=-x_l$, $x=0$ and $x=x_r$.
Applying these conditions to the solutions $\rho_0(x)$ in Eq. (\ref{eqn:SmallfpSol}) shows that the leading order solution is given by the Boltzmann weight, i.e. $\rho_0(x) = \rho_l \exp(-\beta V(x))$ for all $x$. Clearly, this is the correct passive solution. The higher order solutions that follow, i.e. the polarization profile $m_1(x)$ and the density correction $\rho_2(x)$, are plotted in Fig. \ref{fig:SmallfpPlots}. Qualitatively, these plots show the same features as displayed by the numerical solutions in Fig. \ref{fig:TypicalProfiles}: an accumulation of particles facing the barrier at the ratchet sides $x = -x_l$ and $x=x_r$, and a right bulk density $\rho_r$ that exceeds the left bulk density $\rho_l$. To allow for a quantitative comparison, Fig. \ref{fig:SmallfpPlots} also shows polarization profiles $m_x(x)$ and density corrections $\rho(x) - \rho_0(x)$ that were obtained for the 1D RnT model numerically. While the ratchet potential is fixed, with barrier height $\beta V_{\text{max}} = 4$, width $x_l/\ell = 1$, and asymmetry $a=3$, the comparison is made for several degrees of activity, namely $\text{Pe} = 0.1, 0.5, \text{ and } 1$. The analytical and numerical results show good agreement for $\text{Pe} = 0.1$, reasonable agreement for $\text{Pe}=0.5$, and deviate significantly for $\text{Pe} = 1$. All of these observations are as expected, since the analytical solutions (\ref{eqn:SmallfpSol}) are obtained under the assumption $\text{Pe} \ll 1$.
\\
	\indent The most interesting part of solution (\ref{eqn:SmallfpSol}) is the density correction $\rho_2(x)$, as this correction contains the leading order contribution to the difference in bulk densities $\Delta \rho$. To gain some understanding for the meaning of the various terms contributing to $\rho_2(x)$, we point out that for small activity, i.e. for $\text{Pe} \ll 1$, active particles are often understood as passive particles at an effective temperature \cite{CugliandoloEffTemp,Wang2011,MarconiMaggi,Fily2012,Palacci2010,Szamel2014}. In our convention, this effective temperature reads $T_{\text{eff}} = T(1+\text{Pe}^2)$. Therefore, one might think that for our weakly active system the density profile is given by Boltzmann weight at this effective temperature, i.e. by $\rho(x) = A \exp(-V(x)/k_BT_{\text{eff}})$ within one region. Here the prefactor $A$ can depend on the activity $\text{Pe}$. Expanding this effective Boltzmann weight for small $\text{Pe}$ yields the passive solution $\rho_0(x)$, and the terms on the first line of $\rho_2(x)$ in Eq. (\ref{eqn:SmallfpSol}). However, it does \emph{not} reproduce the final two terms that contribute to $\rho_2(x)$ in Eq. (\ref{eqn:SmallfpSol}). Precisely these last two terms are crucial to obtain a nonzero difference $\Delta \rho$ in bulk densities. Indeed, a density profile given solely by the effective Boltzmann weight necessarily yields equal bulk densities $\rho_l = \rho_r$, as the external potential $V(x)$ is equal on either side of the ratchet.\\
\begin{figure}[t] 
  \includegraphics[width=\linewidth]{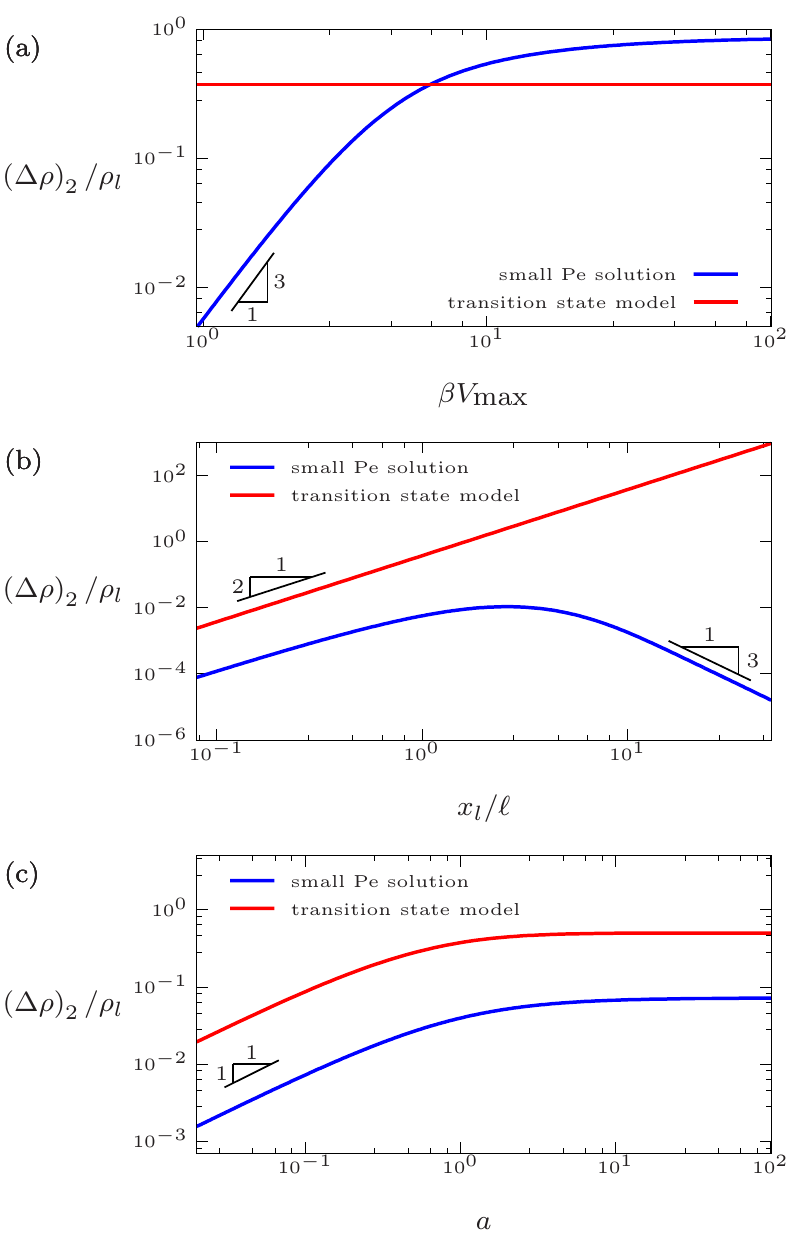}
 \caption{\raggedright Normalized leading order coefficient $(\Delta \rho)_2$ in the expansion of the density difference $\Delta \rho$ for small activity $\text{Pe}$, as found from the analytical $\text{Pe} \ll 1$ solution and as predicted by the transition state model, (a) as a function of the barrier height $\beta V_{\text{max}}$, at fixed barrier width $x_l/\ell = 1$ and asymmetry $a=1$, (b) as a function of the barrier width $x_l/\ell$, at fixed barrier height $\beta V_{\text{max}}=1$ and asymmetry $a=1$, and (c) as a function of the asymmetry $a$, at fixed barrier height $\beta V_{\text{max}} = 2$ and barrier width $x_l/\ell = 1$. The power laws shown by the transition state model in its regime of applicability, i.e. for $\beta V_{\text{max}} \gg 1$ and $x_l/\ell \ll 1$, have exponents that agree with the power laws of the analytical solution. These exponents can be found in Table \ref{tab:powerlaws}. The analytical and transition state solution do not agree quantitavely for these parameter values.} 
 \label{fig:AnaDDs}
\end{figure}
	\indent The analytical expression for the difference in bulk densities $\Delta \rho$, implied by the solutions (\ref{eqn:SmallfpSol}), is rather lengthy and intransparent, and is therefore not shown here. Instead, we show the dependence of $\Delta \rho$ on the activity Pe graphically, in Fig. \ref{fig:DensityDifference}(a), for the same two ratchet potentials as used for the numerical solutions. As the density difference $\Delta \rho$ follows from the correction $\rho_2(x)$, it scales as $\text{Pe}^2$, just like the numerical solutions for Pe $\ll 1$. As shown by Fig. \ref{fig:DensityDifference}(a), the analytical and numerical solutions agree quantitatively up to $\text{Pe} \approx 0.5$, as also found in Fig. \ref{fig:SmallfpPlots}. Before we illustrate how the density difference $\Delta \rho$ depends on the ratchet potential, we extract its dependence on activity $\text{Pe}$ by considering $(\Delta \rho)_2 = \Delta \rho / \text{Pe}^2$, i.e. the leading order coefficient in an expansion of $\Delta \rho$ around $\text{Pe} = 0$.
The coefficient $(\Delta\rho)_2$ is independent of $\text{Pe}$, but still depends on the barrier height $\beta V_{\text{max}}$, the barrier width $x_l/\ell$, and the asymmetry $a$. Its dependence on these ratchet parameters is plotted in Figs. \ref{fig:AnaDDs}(a)-(c), respectively. These figures display all the power law behavior that was obtained numerically in section \ref{sec:NumSol}. The power laws are summarized in Table \ref{tab:powerlaws}.

\section{Transition State Model}
\label{sec:TSModel}

\begin{figure}[t]
 \centering
  \includegraphics[width=\linewidth]{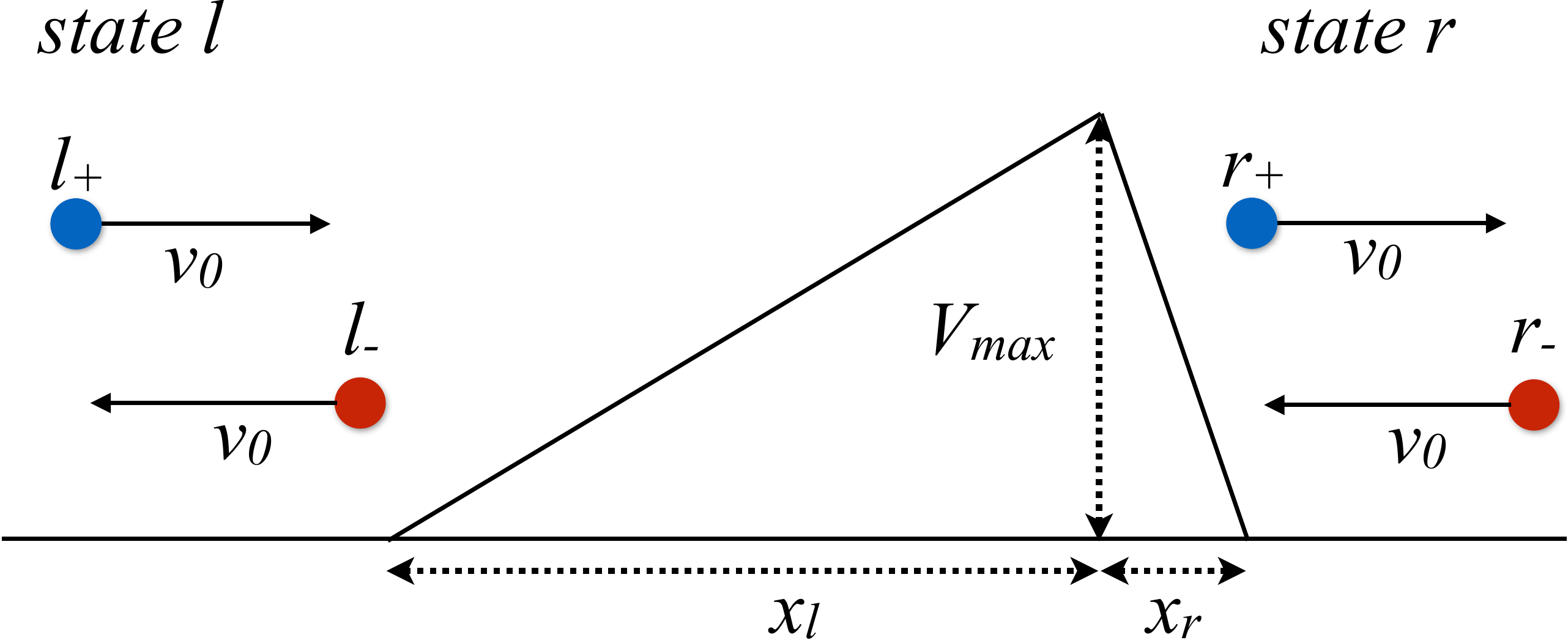}
 \caption{\raggedright Illustration of the states in the transition state model. Particles in the left bulk with positive (negative) $x$-orientation are in state $l_+$($l_-$). Similarly, particles in the right bulk are in state $r_+$ or $r_-$. Within one bulk, particles can change their orientation with rate constant $D_r$. Between the bulks, particles can transition by crossing the potential barrier with the effective Arrhenius rates of Eqs. (\ref{eqn:Rate1}) and (\ref{eqn:OtherRates}), where the effect of self-propulsion is to shift the potential barrier $V_{\text{max}}$ by the work $\gamma v_0 x_l$ ($\gamma v_0 x_r$) performed by the propulsion force when a particles climbs the left (right) slope of the ratchet.}
  \label{fig:TransitionStateModel}
\end{figure}

\noindent As argued in the previous section, the nonzero difference in bulk densities $\Delta \rho$ cannot be accounted for by the effective temperature that is often employed in the weak activity limit. Instead, to understand the behavior of the bulk density difference $\Delta \rho$ better, we propose the following simple transition state model. The model consists of four states, designed to mimic the 1D RnT model in a minimal way. Particles in the bulk to the left of the ratchet, with an orientation in the positive (negative) $x$-direction, are said to be in state $l_+(l_-)$, whereas particles in the bulk to the right of the ratchet, with positive (negative) $x$-orientation, are in state $r_+(r_-)$. This setting is illustrated in Fig. \ref{fig:TransitionStateModel}. Particles can change their orientation, i.e. transition from $l_{\pm}$ to $l_{\mp}$, and from $r_{\pm}$ to $r_{\mp}$, with a rate $D_r$. Furthermore, particles can cross the potential barrier and transition between the $l$- and $r$-states. The associated rate constants are assumed to be given by modified Arrhenius rates \cite{Friddle2008,Eyring1943,Bell1978}, where the effect of self-propulsion is to effectively increase or decrease the potential barrier. For example, the rate to transition from $l_+$ to $r_+$ is 
\begin{align*}
 \label{eqn:Rate1}
  k_{l_+ \rightarrow r_+}
  =
  \frac{\nu_l}{L_l}\exp\left[-\beta(V_{\text{max}}-\gamma v_0 x_l)\right].
 \numberthis
\end{align*}
\noindent As the propulsion force helps the particle to cross the barrier, it effectively lowers the potential barrier $V_{\text{max}}$ by the work $\gamma v_0 x_l$ that the propulsion force performs when the particle climbs the left slope of the ratchet. This modified Arrhenius rate is expected to be valid under the assumptions (a) of a large barrier height $\beta V_{\text{max}} \gg 1$, which is a condition for the Arrhenius rates to be valid even for passive systems \cite{Kramers1940}, (b) of a ratchet potential that is typically crossed faster than a particle reorients, which can be achieved by making the barrier width $x_l/\ell$ sufficiently small, and (c) that the work $\gamma v_0 x_l$ performed by the propulsion force is much smaller than the barrier height $V_{\text{max}}$. We point out that assumption (c) can be rewritten as $\text{Pe} \ll \beta V_{\text{max}} \ell / x_l$. This means that if assumptions (a) and (b) are satisfied, which imply that $\beta V_{\text{max}} \ell / x_l \gg 1$, then assumption (c) is not much further restrictive on the activity $\text{Pe}$. The remaining rate constants follow along a similar reasoning as
\begin{align*}
 \label{eqn:OtherRates}
 \begin{alignedat}{1}
   k_{l_- \rightarrow r_-}
  &=
  \frac{\nu_l}{L_l}\exp\left[-\beta(V_{\text{max}}+\gamma v_0 x_l)\right], \\   
   k_{r_+ \rightarrow l_+}
  &=
  \frac{\nu_r}{L_r}\exp\left[-\beta(V_{\text{max}}+\gamma v_0 x_r)\right], \\   
   k_{r_- \rightarrow l_-}
  &=
  \frac{\nu_r}{L_r}\exp\left[-\beta(V_{\text{max}}-\gamma v_0 x_r)\right].
 \end{alignedat}
 \numberthis
\end{align*}
\noindent For large bulks on either side of the ratchet, the attempt frequencies in the rate expressions (\ref{eqn:Rate1}) and (\ref{eqn:OtherRates}) are inversely proportional to the size of the bulk that is being transitioned from. This size is denoted by $L_l$ for the left bulk, and by $L_r$ for the right bulk. Therefore, the factors $\nu_l$ and $\nu_r$ are independent of the bulk sizes $L_l$ and $L_r$, and can only depend on the shape of the rachet potential, i.e. on its height $\beta V_{\text{max}}$, on its width $x_l/\ell$, and on its asymmetry $a$.
\\
	\indent We denote the number of particles in the $l_{\pm}$ and $r_{\pm}$ states by $N_{l_{\pm}}(t)$ and $N_{r_{\pm}}(t)$, respectively. The time evolution of these particle numbers follows from the rates outlined above. For example, the number of particles $N_{l_+}(t)$ in state $l_+$ evolves according to the rate equation 
\begin{flalign*}
 \label{eqn:RateEq}
  \partial_t N_{l_+}
  \! = \!
  - \! \left(D_r+k_{l_+ \rightarrow r_+} \! \right) \! N_{l_+}
  +D_r N_{l_-} +
  \mathrlap{k_{r_+ \rightarrow l_+} \! N_{r_+} \! .}
 \numberthis &&
\end{flalign*}
\noindent Similar equations hold for the particle numbers $N_{l_{-}}(t)$, $N_{r_{+}}(t)$ and $N_{r_{-}}(t)$. These rate equations can be solved in steady state, i.e. when $\partial_t N_{l_{\pm}} = \partial_t N_{r_{\pm}} = 0$, for the particle numbers $N_{l_{\pm}}$ and $N_{r_{\pm}}$. We consider infinitely large bulks, i.e. $L_l,L_r \rightarrow \infty$. In this case, the solutions show that $N_{l_+} = N_{l_-}$ and $N_{r_+} = N_{r_-}$, such that the $l$ and $r$ states correspond to isotropic bulks. Furthermore, the solution shows that the bulk densities $\rho_l = (N_{l_+} + N_{l_-})/L_l$ and $\rho_r = (N_{r_+} + N_{r_-})/L_r$ differ by an amount $\Delta \rho = \rho_r - \rho_l$ given by
\begin{align*}
 \label{eqn:TwoStateSol}
  \frac{\Delta \rho}{\rho_l}
  =
  \frac{\nu_l}{\nu_r}
  \frac{\cosh\left(\text{Pe} \, x_l/\ell\right)-\cosh\left(\text{Pe} \, x_r/\ell\right)}
       {\cosh\left(\text{Pe} \, x_r/\ell\right)},
 \numberthis
\end{align*}
\noindent where we recall that $x_r = (1+a)^{-1} x_l$. We point out that the ratio $\nu_l/\nu_r$ can generally depend on the ratchet parameters $\beta V_{\text{max}}$, $x_l/\ell$, and $a$. However, in the following we simply assume $\nu_l/\nu_r = 1$, which is justified for nearly symmetric ratchets.
\\
\indent To enable a comparison with the analytical solution of the previous section, we now focus on the limit of weak activity, i.e. of $\text{Pe} \ll 1$. This ensures assumption (c) to be satisfied, but we emphasize that the transition state model is not limited to weak activity. We expand the density difference (\ref{eqn:TwoStateSol}) as $\Delta \rho = (\Delta \rho)_2 \ \text{Pe}^2 + \Ov(\text{Pe}^4)$, and compare the coefficient $\left(\Delta\rho\right)_2$ with the same coefficient obtained in section \ref{sec:NumSol} for the analytical solution in the weak activity limit. The coefficient $(\Delta\rho)_2$ is plotted in Figs. \ref{fig:AnaDDs}(a)-(c), as a function of the of the barrier height $\beta V_{\text{max}}$, the barrier width $x_l/\ell$, and the barrier asymmetry $a$, respectively. Fig. \ref{fig:AnaDDs}(a) merely illustrates that the density difference (\ref{eqn:TwoStateSol}) is independent of the barrier height $\beta V_{\text{max}}$. This independency agrees with the asymptotic behavior displayed by the analytical solution for large barrier heights $\beta V_{\text{max}} \gg 1$. Note that the regime $\beta V_{\text{max}} \gg 1$ is indeed assumed for the modified Arrhenius rates (assumption (b)). Fig. \ref{fig:AnaDDs}(b) illustrates that the density difference predicted by the transition state model scales quadratically with the barrier width, i.e. that $\Delta \rho \propto (x_l/\ell)^2$. This scaling agrees with the scaling of the analytical solution for the regime of small barrier widths $x_l/\ell \ll 1$. Again, this regime is assumed for the modified Arrhenius rates, as having a small barrier width is required for having particles cross the ratchet faster than they typically reorient (assumption (c)). Finally, Fig. \ref{fig:AnaDDs}(c) illustrates that the density difference predicted by the transition state model scales linearly with the barrier asymmetry for nearly symmetric ratchets, i.e. $\Delta \rho \propto a$ for $a\ll1$, and asymptotically for very asymmetric ratchets, i.e. $\Delta \rho \propto a^0$ for $a\gg1$. Both scalings are also displayed by the analytical solution. All these power laws can again be found in Table \ref{tab:powerlaws}.\\ 
	\indent Of course, the transition state model reproduces only the power laws that lie inside its regime of applicability. However, the fact this simple model \emph{does} reproduce all these power laws is quite remarkable, since, as discussed in section \ref{sec:NumSol}, most of these scalings are nontrivial. Furthermore, we note that the transition state model can also be solved for finite bulk sizes, which in fact predicts a turnover of the density difference $\Delta \rho$ as a function of activity Pe, as observed in Fig. \ref{fig:DensityDifference}(a).\\
	\indent Quantitatively, Fig. \ref{fig:AnaDDs} clearly shows that the predictions of the transition state model typically differ from the analytical solution by an order of magnitude. A possible reason for this disagreement is that these plots are made for parameters values that do not satisfy assumptions (a) and (b) that underly the modified Arrhenius rates. In fact, it turned out to be impossible to satisfy these assumptions simultaneously with feasible parameter values. The root of the difficulty is that the time it takes a particle to cross the potential barrier increases with the barrier height $\beta V_{\text{max}}$. As a consequence, having a barrier that is simultaneously very high (assumption (a)), and typically crossed faster than a particle reorients (assumption (b)), turns out to require unrealistically small barrier widths $x_l/\ell$.
The quantitative mismatch of the transition state model with the full solution for small activity might also be attributed to the assumption that the prefactors $\nu_l$ and $\nu_r$ in the rate expressions (\ref{eqn:Rate1}) and (\ref{eqn:OtherRates}) are not exactly identical, but in fact might depend on the precise shape of the barrier. However, this possibility goes beyond the current scope of this paper, and we leave it for future study.\\
	\indent We conclude that, whereas it was not possible to test the predictions of the transition state model in its regime of applicability quantitatively, the model \emph{does} reproduce the complete power law behavior of this regime correctly.

\section{Discussion}
\label{sec:Discussion}

\noindent The most interesting aspect of the studied system is that the external potential has a long-range influence on the density profile. This is in sharp contrast to an ideal gas in equilibrium, whose density profile is only a function of the \emph{local} external potential. So what ingredients are necessary to obtain this effect? To answer this question, we consider the 1D RnT model subject to a general external potential $V(x)$. Furthermore, we introduce the particle current $J(x)$ and the orientation current $J_m(x)$ that appear in the evolution equations (\ref{eqn:RnTEqs}), i.e.
\begin{align*}
 \label{eqn:1DCurrents} 
 \begin{alignedat}{1}
  J(x) &=
    \sqrt{2} v_0 m_x   
    -\frac{1}{\gamma}(\partial_xV)\rho
    -D_t\partial_x\rho,\\
  J_m(x) &=
    \frac{v_0}{\sqrt{2}} \rho   
    - \frac{1}{\gamma}(\partial_xV)m_x
    - D_t \partial_x m_x.
 \end{alignedat}
 \numberthis
\end{align*}
\noindent We focus on a state that is steady, such that $J(x) = \text{constant} \equiv J$, and flux-free, such that $J = 0$. Then Eqs. (\ref{eqn:RnTEqs}) and (\ref{eqn:1DCurrents}) can be recast into the first order differential equation
\begin{align*}
 \label{eqn:DiffEq}
  \ell \partial_x \mb{Y}(x)
  =
  \mbs{\mathcal{M}}(x)
  \mb{Y}(x)
 \numberthis
\end{align*}
\noindent 
for the three (non-dimensionalized) unknowns $\mb{Y}(x) \equiv \left( \ell \rho(x), \ell m_x(x), J_m(x)/D_r\right)^T$. The coefficient matrix in Eq. (\ref{eqn:DiffEq}) is given by 
\begin{align*}
 \label{eqn:M}
  \mbs{\mathcal{M}}(x)
  =
  \begin{bmatrix}
   f(x)               & \sqrt{2}\text{Pe} & 0  \\
   \text{Pe}/\sqrt{2} & f(x)              & -1 \\
   0                  & -1                & 0
  \end{bmatrix},
 \numberthis
\end{align*}
\noindent where $f(x) \equiv -\beta \ell \partial_x V(x)$ is the dimensionless external force, that is now a function of position $x$. For a passive system (Pe $=0$), Eqs. (\ref{eqn:DiffEq}) and (\ref{eqn:M}) show that the density equation decouples. In this case, the density profile is solved by the Boltzmann weight, i.e. $\rho(x) \propto \exp(-\beta V(x))$, as required in thermodynamic equilibrium. For the general case, we observe that, \emph{if} the coefficient matrix $\mbs{\mathcal{M}}(x)$ commutes with its integral $\int_{x_0}^x \mathrm{d}x'\mbs{\mathcal{M}}(x')$, \emph{then} Eq. (\ref{eqn:DiffEq}) is solved by
\begin{align*}
 \label{eqn:SpecialSolution}
  \mb{Y}(x)
  = 
  \exp\left(\frac{1}{\ell}\int_{x_0}^x\mathrm{d}x'\mbs{\mathcal{M}}(x')\right)
  \cdot
  \begin{pmatrix}
   c_1 \\ c_2 \\ c_3
  \end{pmatrix},
 \numberthis
\end{align*}
\noindent where the integration constants $c_1$, $c_2$ and $c_3$ are to be determined from boundary conditions.
\noindent Here $x_0$ is an arbitrary reference position. By virtue of $\int_{x_0}^x\mathrm{d}x'f(x')=-\beta \ell V(x)$, the solution (\ref{eqn:SpecialSolution}) \emph{is} a local function of the external potential. An explicit calculation of the commutator shows that $[ \mbs{\mathcal{M}}(x),\int_{x_0}^x\mathrm{d}x'\mbs{\mathcal{M}}(x')] = 0$ if and only if $\beta (V(x)-V(x_0)) = -f(x) \, (x-x_0)/\ell$, i.e. if the external potential is a linear function of $x$. Therefore, for linear potentials, the density profile \emph{is} a local function of the external potential. This explains why in a gravitational field the density profile \emph{can} be found as a local function of the external potential, and why sedimentation profiles stand a chance to be described in terms of an effective temperature in the first place\cite{Tailleur2008,Tailleur2009,SelfPumpingState1,Palacci2010,ES,Wolff2013,Szamel2014,ABPvsRnT,EoSExperiment,Stark2016,MatthiasSedimentation,Sedimentation2018}. However, for nonlinear external potentials, e.g. for the ratchet studied here that is only piecewise linear, the solution (\ref{eqn:SpecialSolution}) is \emph{not} valid, and a nonlocal dependence on the external potential is to be expected. Therefore, for the ratchet potential (\ref{eqn:RatchetPotential}), the kinks at $x=-x_l$, $x=0$ and $x=x_r$ are crucial to have a density that depends nonlocally on the external potential. Indeed, in the analytical solution for weak activity, presented in section \ref{sec:SmallfpSol}, the nonlocal dependence of the right bulk density $\rho_r$ on the external potential enters through the fact that the integration constants in Eq. (\ref{eqn:SmallfpSol}) are found from continuity conditions that are applied precisely at the positions of these kinks.\\ 
	\indent Summarizing, in order to have the external potential influence the steady-state density of ideal particles in a nonlocal way, one needs to have (1) particles that are active (such that the system is out of thermodynamic equilibrium), and (2) an external potential that is nonlinear. Thereby, the 1D RnT particles in the ratchet potential (\ref{eqn:RatchetPotential}) illustrate the nonlocal, and even long-range, influence of the external potential in a most minimal way.\\
	\indent In the discussion above, we have only shown that a linear external potential yields a density profile that is a strictly local function of the potential. Thereby, a nonlinear potential is not guaranteed to influence the density (arbitrarily) far away, and indeed other criteria have been discussed in the literature. For example, in the context of active Ornstein-Uhlenbeck particles, approximate locality was shown for a wide class of nonlinear potentials \cite{MarconiMaggi,MarconiMaggi2}, and it was argued that in order to lose this property it is crucial to have an external potential with nonconvex regions \cite{Fily2017b}. More generally, the fact that the potential barrier is more easily crossed from one side than from the other is a rectification effect, and it has been shown that such effects can occur when the dynamics break time-reversal symmetry, while also the spatial mirror symmetry is broken \cite{Magnasco1993,Prost1994}. In our case, these criteria are met by the presence of activity, and by having a ratchet that is asymmetric ($a\neq 0$), respectively.\\
	\indent Our results are also fully consistent with the work by Baek et al. \cite{Yongjoo2018}, who study the effect of placing a nonspherical body in a two-dimensional fluid of ABPs. They show that such an inclusion leads to a steady state with a density perturbation that scales in the far field as $1/r$, where $r$ is the distance to the body. Repeating their derivation for the 1D RnT model in our setting yields a far-field density perturbation that is simply constant, i.e. independent of $r$. This is consistent with our findings. Furthermore, under suitable conditions, in particular that the external potential is small everywhere, the authors of \cite{Yongjoo2018} derive that the far-field density perturbation scales as $(V_{\text{max}})^3$. This confirms our finding of the powerlaw $\Delta \rho \propto (\beta V_{\text{max}})^3$ for small potential barriers $\beta V_{\text{max}} \ll 1$. Moreover, it suggests that this scaling is not limited to the sawtooth-shaped potential barrier considered here, but also holds for external potentials of more general shape.

\section{Conclusions}
\label{sec:Conclusion}
\noindent We have studied the distribution of noninteracting, active particles over two bulks separated by a ratchet potential. The active particles were modelled both as two-dimensional ABPs, and as one-dimensional RnT particles. Our numerical solutions to the steady state Smoluchowski equations show that the ratchet potential influences the distribution of particles over the bulks, even though the potential is short-ranged itself. Thus, the external potential exerts a long-range influence on the density profile. We have shown that such a (highly) nonlocal influence can occur for noninteracting particles only when they are (1) active, and (2) subject to an external potential that is nonlinear. Thereby, the piecewise linear setup considered in this work captures this long-range influence in a most minimal way.
\\
	\indent To characterize the influence of the external potential, we have described how the difference in bulk densities depends on activity, as well as on the ratchet potential itself. Both models of active particles showed consistent power law behavior that is summarized in Table \ref{tab:powerlaws}.\\
	\indent To understand the long-range influence of the potential in the simplest case possible, we focussed on the limit of weak activity. While weakly active systems are often described by an effective temperature, our analytical solution explicitly shows that the long-range influence of the ratchet potential cannot be rationalized in this way. Instead, we propose a simple transition state model, in which particles can cross the potential barrier by Arrhenius rates with an effective barrier height that depends on the degree of activity. While the model could not be tested quantitatively, as its underlying assumptions could not be simultaneously satisfied for feasible parameter values, it does reproduce - in its regime of applicability - the complete power law behavior of the distribution of particles over the bulks.\\
	\indent Future questions are whether the power law behavior can be understood also outside the regime where the transition state model applies, and whether the power laws also hold for potential barriers of more generic shape than the sawtooth of Fig. \ref{fig:RatchetPotential}. 
Our work illustrates that even weakly active, noninteracting particles pose challenges that are fundamental to nonequilibrium systems,
and, moreover,  that an external potential can exert a long-range influence in such systems.
We expect that incorporating such long-range and nonlocal effects will be part of a more generic statistical mechanical description of nonequilibrium systems. 

\section{Acknowledgments}
\noindent This work is part of the D-ITP consortium, a program of the Netherlands Organisation for Scientific Research (NWO) that is funded by the Dutch Ministry of Education, Culture and Science (OCW). We acknowledge funding of a NWO-VICI grant. S.P. and M.D. acknowledge the funding from the Industrial Partnership Programme `Computational Sciences for Energy Research' (Grant No. 14CSER020) of the Foundation for Fundamental Research on Matter (FOM), which is part of the Netherlands Organization for Scientific Research (NWO). This research programme is co-financed by Shell Global Solutions International B.V.

\appendix*
\section{Weak Activity Solution}
\noindent In this appendix, we derive the analytical solutions (\ref{eqn:SmallfpSol}), i.e. the steady state solutions to the 1D RnT Eqs. (\ref{eqn:RnTEqs}) in the limit of weak activity. First, we define the particle flux $J(x) \equiv \sqrt{2} v_0 m_x - \gamma^{-1} (\partial_x V)\rho - D_t \partial_x\rho$, such that the evolution equation for the density, given by Eq. (\ref{eqn:RnTEqs}), reads $\partial_t \rho = -\partial_x J(x)$. Having a steady state ($\partial_t \rho = 0$) implies that $J(x) \equiv J$ is constant, i.e. independent of $x$. The boundary condition of having a bulk at $x = x_{\text{max}}$ that is homogeneous and isotropic, and hence flux-free, then implies $J=0$. The equation $J(x)=0$ has to be solved together with the steady state condition for the polarization implied by Eq. (\ref{eqn:RnTEqs}). In dimensionless form, these equations read
\begin{align*}
 \label{eqn:DimlessRnTEqs}
 \begin{alignedat}{1}
  0 &= \sqrt{2}\text{Pe} \, m_x + f(x) \rho - \ell \partial_x\rho, \\
  0 &= -\ell\partial_x 
         \left\{
          \frac{\text{Pe}}{\sqrt{2}} \rho
          +f(x) m_x
          -\ell\partial_xm_x
         \right\}
        -m_x.
 \end{alignedat}
 \numberthis
\end{align*}
Here, we defined the non-dimensionalized external force $f(x) \equiv -\beta\ell\partial_xV(x)$. We shall solve Eqs. (\ref{eqn:DimlessRnTEqs}) separately for every region where the ratchet potential (\ref{eqn:RatchetPotential}) is a linear function. Within one such region, $f(x) = f$ is constant, namely $f=0$ to the left and to the right of the potential barrier, $f=-\beta V_{\text{max}} \ell/x_l$ on the left slope of the barrier, and $f=-\beta V_{\text{max}} \ell/x_r$ on the right slope. We treat these cases simultaneously by simply writing $f(x) = f$, and keeping in mind that the solution holds only within one region. Furthermore, we focus on the limit of weak activity, i.e. of $\text{Pe} \ll 1$, and expand the density as $\rho(x) = \rho_0(x) + \text{Pe}^2 \rho_2(x) + \Ov(\text{Pe}^4)$, and the polarization as $m_x(x) = \text{Pe} \, m_1(x) + \Ov(\text{Pe}^3)$, as explained in the main text. We insert these expansions into Eqs. (\ref{eqn:DimlessRnTEqs}), and solve order by order in $\text{Pe}$. To zeroth order in $\text{Pe}$, the equations read $f \rho_0 - \ell\partial_x \rho_0 = 0$, and are solved by
\begin{align*}
 \label{eqn:Pe0}
  \rho_0(x)=A_0 e^{fx/\ell},
 \numberthis
\end{align*}
\noindent where $A_0$ is an integration constant. Note that Eq. (\ref{eqn:Pe0}) is the Boltzmann weight, and hence the correct passive solution for noninteracting particles. To linear order in $\text{Pe}$, the equations read
\begin{align*}
 \label{eqn:Pe1Eq}
  -\ell\partial_x
    \left(
     f m_1 - \ell \partial_xm_1
    \right)
   -m_1
   =
   \frac{1}{\sqrt{2}} \ell \partial_x \rho_0,
 \numberthis
\end{align*}
\noindent where $\rho_0(x)$ is given by Eq. (\ref{eqn:Pe0}). The solution to Eq. (\ref{eqn:Pe1Eq}) is
\begin{align*}
 \label{eqn:Pe1Sol}
  m_1(x)
  =
  -\frac{A_0}{\sqrt{2}} f e^{fx/\ell}
  +B_+ e^{c_+x/\ell}
  +B_- e^{c_-x/\ell},
 \numberthis
\end{align*}
\noindent where $B_+$ and $B_-$ are integration constants, and where $c_{\pm} \equiv (f \pm \sqrt{f^2+4})/2$. To quadratic order in $\text{Pe}$, the equations read
\begin{align*}
 \label{eqn:Pe2Eq}
  f \rho_2 - \ell \partial_x \rho_2
  =
  -\sqrt{2} m_1,
 \numberthis
\end{align*}
\noindent where $m_1(x)$ is given by Eq. (\ref{eqn:Pe1Sol}). The solution to Eq. (\ref{eqn:Pe2Eq}) is given by
\begin{align*}
 \label{eqn:Pe2Sol}
 \begin{alignedat}{1}
  \rho_2(x)
  =
  \left[A_2-A_0 f x/\ell\right]e^{fx/\ell}
  &+ \frac{\sqrt{2}B_+}{c_+ - f} e^{c_+ x/\ell} \\
  &+ \frac{\sqrt{2}B_-}{c_- - f} e^{c_- x/\ell},
 \end{alignedat}
 \numberthis
\end{align*}
\noindent where $A_2$ is another integration constant. Together, Eqs. (\ref{eqn:Pe0}), (\ref{eqn:Pe1Sol}) and (\ref{eqn:Pe2Sol}) constitute the solution (\ref{eqn:SmallfpSol}) of the main text.\\
	\indent As emphasized above, these solutions hold within every region separately. Therefore, the values of the integration constants $A_0$, $A_2$, $B_+$, and $B_-$ can differ per region. These values are determined from the boundary conditions outlined in section \ref{subsec:NumSolA}, i.e. $\rho(-\infty)=\rho_l$, $m_x(-\infty)=0$, and $m_x(\infty)=0$ (we take $x_{\text{res}} \rightarrow -\infty$ and $x_{\text{max}} \rightarrow \infty$), and from the requirements that the density $\rho(x)$, the polarization $m_x(x)$, and the orientation flux $J_m(x) \equiv \text{Pe} \, \rho/\sqrt{2} +f(x)m_x - \ell\partial_xm_x$ all be continuous at the region boundaries $x=-x_l$, $x=0$ and $x=x_r$. The conditions that $\rho(-\infty) = \rho_l$ and that the the density $\rho(x)$ be continuous straightforwardly imply that $A_0 = \rho_l$ everywhere. However, the values that follow for the other integration constants $A_2$, $B_+$, and $B_-$ are mostly lengthy, and intransparent, and therefore not shown. The same is true for the leading order difference in bulk densities $\Delta \rho = \rho_r - \rho_l = \text{Pe}^2 A_2|_{x>x_r}$, whose dependence on the parameters of the problem is instead depicted graphically, in Figs. \ref{fig:DensityDifference}(a) and \ref{fig:AnaDDs}(a)-(c).


\end{document}